\documentclass[pra,showpacs,superscriptaddress,twocolumn]{revtex4-1}
\usepackage{amsfonts}
\usepackage{amssymb}
\usepackage{amsmath}
\usepackage{graphicx}
\usepackage{epsfig}
\begin{document}

\title{Topological phases and edge states in a non-Hermitian trimerized optical lattice}
\author{L. Jin}
\email{jinliang@nankai.edu.cn}
\affiliation{School of Physics, Nankai University, Tianjin 300071, China}

\begin{abstract}
Topologically engineered optical materials support robust light transport.
Herein, the investigated non-Hermitian lattice is
trimerized and inhomogeneously coupled using uniform intracell coupling. The
topological properties of the coupled waveguide lattice are evaluated, the $%
\mathcal{PT}$-symmetric phase of a $\mathcal{PT}$-symmetric lattice can have different topologies; the edge states depend on the lattice size, boundary configuration, and competition between the coupling and degree of non-Hermiticity. The topologically nontrivial region
extends in the presence of periodic gain and loss. The nonzero geometric
phases accumulated by the Bloch bands indicate the existence of
topologically protected edge states between the band gaps. The
unidirectional amplification and attenuation zero modes appear above a
threshold degree of non-Hermiticity, which facilitate the development of a
robust optical diode.
\end{abstract}

\maketitle

%11.30.Er, Charge conjugation, parity, time reversal, and other discrete symmetries
%42.60.Da, Resonators, cavities, amplifiers, arrays, and rings
%03.65.Vf, Phases: geometric; dynamic or topological

\section{Introduction}

Topological insulators are novel states of solid-state materials that have
an insulating bulk band gap and a conducting edge or surface~\cite{RMP10}.
The edge or surface states are symmetry-protected against local disorder,
and valuable for quantum transport and computation in quantum (anomalous)
Hall insulators and quantum spin Hall insulators~\cite{RMP11}. The optical
realization of topological systems has stimulated the field of topological
photonics, which enables the experimental studies of topological systems that are difficult to realize in condensed matter physics. The photons in
coupled waveguides and optical lattices are manipulated in a manner similar
to the electrons in solids, providing intriguing opportunities for novel
optical devices~\cite{Khanikaev,Rechtsman,TopoPhot}. The topologically
protected unidirectional interface state propagates robustly against local
impurities. This was experimentally demonstrated in coupled waveguide ring
resonators~\cite{Hafezi}. Synthesizing artificial gauge fields for ultracold
atoms in optical lattices enables the construction of a two-dimensional
topological system~\cite{Zollor}. The Su-Schrieffer-Heeger (SSH) model~\cite%
{SSHPRL} is the simplest system that has topologically nontrivial edge
states~\cite%
{Ryu,KM,BernevigPRL,Bernevig,Qi,Fu,Esaki,Kitaev,RyuNJP,SChenPRL,GuoHM,WangHSciBulletin}%
; this model has been realized for a dimerized optical superlattice, wherein
the topological properties of Bloch bands were experimentally measured~\cite%
{IBloch}.

As progress on topological photonics has advanced, tremendous interest has
also been paid to parity-time ($\mathcal{PT}$) symmetric non-Hermitian systems in coupled
optical waveguides~\cite{Muga,OL,Musslimani,AGuo}, resonators~\cite%
{Jing,PengNP,PengScience,Chang,JingSR,XYL,ZPL}, atoms and atomic lattices~%
\cite{CHang,ZZhang}. $\mathcal{PT}$-symmetric systems can have an entirely
real spectrum although they are non-Hermitian~\cite%
{Bender,Dorey,Ali,Jones,Klaiman,Znojil,LJin09,Rotter09,YNJ}.\ Intriguing
phenomena have been revealed including fast evolution~\cite{Faster,AliPRL07,Guenther,FasterObs}, power oscillation~\cite{CERuter}, unidirectional reflectionless~\cite{Feng}, and coherent absorption~\cite{CPAChong,CPAScience,CPAHChen}. The $\mathcal{%
PT}$-symmetric properties of an SSH chain with a pair of $\mathcal{PT}$-symmetric potentials at boundaries have been studied~\cite{SChen,LS,Ruzicka}; the $\mathcal{PT}$ transition threshold has a power law decay with the SSH chain size~\cite{LS}; Hilbert space inner product has been constructed in the framework of pseudo-Hermitian quantum mechanics, which provides deep insight and novel physics on the non-Hermitian topological systems~\cite{Ruzicka}. Moreover, the edge states are unaffected by the gain and loss when zero
probabilities are located at the edges~\cite{Yuce15,JLSSH}.

Because extensive progress has been made in topological photonics and $\mathcal{PT}$-symmetric optics, optical analysis of topological systems has
been extended to non-Hermitian systems~\cite{Chong,Zeuner,Malzard,Schomerus}. The
topological interface states in non-Hermitian systems have been
systematically discussed~\cite{Leykam}. Topologically protected $%
\mathcal{PT}$-symmetric interface states were demonstrated in coupled
resonators~\cite{Szameit}. Robust light interface states were discovered at
the interface between SSH chains with distinct non-Hermiticity~\cite{FengLSR}.
Topological properties are characterized using the generalized winding
number in non-Hermitian systems~\cite{Esaki2011}. $\mathcal{PT}$-symmetric
non-Hermitian Aubry-Andr\'{e} systems~\cite{Yuce14,YNJ2016} and Kitaev models~\cite{PTKitaev,Wunner} were theoretically investigated. Evidence reveals that universal non-Hermiticity may alter topological regions~\cite{Hughes2011}, but topological properties are robust against
local non-Hermiticity.

In this paper, we investigate a coupled waveguide lattice that has a
trimerized unit cell and in which gain and loss are balanced and separated
by a passive waveguide. The lattice has universal non-Hermiticity. The intracell coupling, $g_{1}$, is uniform and
differs from the intercell coupling, $g_{2}$. The nontrivial topology of the
lattice implies that it has a conducting edge or surface. Under the
periodical boundary condition, the geometric phases accumulated are $\pi $
when the upper and lower bands circle one loop in the Brillouin zone in the
topologically nontrivial region; corresponding edge states exist between the
band gaps under an open boundary condition. In $\mathcal{PT}$-symmetric configurations, the $\mathcal{PT}$-symmetric
regions possess different topologies depending on the coupling configuration
and universal non-Hermiticity. In particular, we discover a non-Hermitian
threshold above which there exist unidirectional amplified and damped edge
states, which are located at opposite boundaries with identical resonant frequency. The
topologically nontrivial regions with edge states are widened because of the
non-Hermitian periodic gain and loss. The threshold corresponds to a
bifurcation point in the imaginary part of the spectrum. The topological
region expands when the lattice has substantial non-Hermiticity.

The reminder of the paper is organized as follows. In Sec.~\ref{II}, we
discuss the topological properties of a Hermitian trimerized lattice. In
Secs.~\ref{III} and~\ref{IV}, we investigate the lattice's $\mathcal{PT}$%
-symmetric non-Hermitian extension based on coupled waveguides. The band
structure, $\mathcal{PT}$-symmetric phases, and edge states are
investigated. The results are summarized in Sec.~\ref{V}.

\section{The trimerized lattice}

\label{II}

\begin{figure}[tb]
\includegraphics[bb=0 0 560 380, width=8.6 cm, clip]{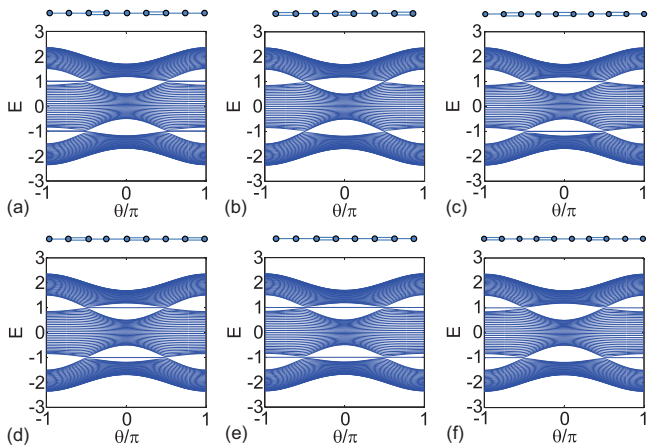}
\caption{(Color online) Energy bands for (a)-(c) $\mathcal{PT}$-symmetric
and (d)-(f) non-$\mathcal{PT}$-symmetric configurations. (a),(d) $N=90$; (b),(e)
$N=89$; and (c),(f) $N=88$. Lattice configurations are displayed above energy bands. $\Delta =1/2$.}
\label{fig1}
\end{figure}

We focus on the topological properties of a trimerized lattice, which has
more edge states than a dimerized SSH chain~\cite{SChen,LS,Ruzicka}. The topologically protected
edge states are related to the lattice structure and determined by the
boundary~\cite{AAHPRL,YNJAA}. The trimerized lattice investigated herein
comprises three groups, which are indicated by lattice number $N=3n$, $3n-1$%
, and $3n-2$; when considering the coupling configurations, the lattice is
divided into three groups by their inhomogeneous couplings at the boundary: $%
g_{1}$-$g_{1}$-$g_{2}$-, $g_{2}$-$g_{1}$-$g_{1}$-, and $g_{1}$-$g_{2}$-$%
g_{1} $-. The lattice has only six topologically distinct configurations because of
its reflection symmetry. These configurations are illustrated above their spectra presented in Fig.~\ref{fig1}, where each site represents a
waveguide. The three configurations in Fig.~\ref{fig1}(a)-\ref{fig1}(c) are $%
\mathcal{PT}$-symmetric, whereas those in Fig.~\ref{fig1}(d)-\ref{fig1}(f)
are not. The differences in the configuration of the trimerized lattice
result in more edge states in the topologically nontrivial phase. The
trimerized lattice that we consider can be modeled using a one-dimensional
off-diagonal AAH Hamiltonian $H_{\mathrm{AAH}}=\sum_{j}\left[ 1+\lambda \cos
\left( 2\pi \beta j+\phi _{\lambda }\right) \right] a_{j}^{\dagger }a_{j+1}+%
\mathrm{H.c.} $, where $a_{j}^{\dagger }$ ($a_{j}$) is the creation
(annihilation) operator for bosonic particles. A rational number of $\beta
=1/3$ leads to a trimerized lattice with three bands. The amplitude $\lambda
$ and phase factor $\phi _{\lambda }$ control the modulation of coupling
strength. At $\phi _{\lambda }=2m\pi /3$ (where $m$ is an integer), the
three repeated couplings are $\{g_{1},g_{1},g_{2}\}$ with $g_{1}=1-\lambda
/2 $ and $g_{2}=1+\lambda $.

\begin{figure}[tb]
\includegraphics[bb=0 0 550 180, width=8.8 cm, clip]{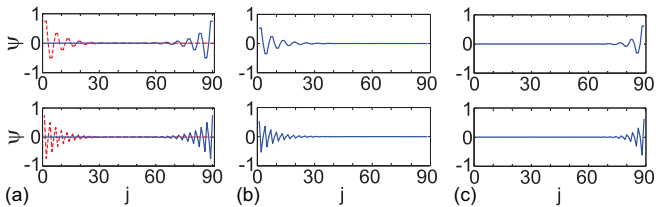}
\caption{(Color online) Three types of edge states for configurations in (a) Fig.~\protect\ref{fig1}(a) at $\protect\theta =\protect\pi $, (b) Fig.~\protect\ref{fig1}(e) at $\protect\theta =\protect\pi $, and (c) Fig.~\protect\ref{fig1}(e) at $\protect\theta =0$. The edge states in Fig.~\protect\ref{fig1}(c) at
$\protect\theta =0$ are depicted in (a); those in Fig.~\protect\ref{fig1}(d) at $\protect\theta =0$ are depicted in (b); and those in Fig.~\protect\ref{fig1}(f) at $\protect\theta =\protect\pi $ are  depicted in (c). $\Delta =1/2$.}
\label{fig2}
\end{figure}

To investigate the topological properties of the trimerized lattice, we set
the intracell coupling $g_{1}$ to unity and force the intercell coupling to
change periodically as
\begin{equation}
g_{2}=1-\Delta \cos \theta .
\end{equation}%
The energy bands are plotted as a function of $\theta $ in Fig.~\ref{fig1}.
The single line indicates the intracell coupling $g_{1}$, and the double
lines indicate the intercell coupling $g_{2}$. Edge states with energies $%
\pm 1$ exist between the upper and lower band gaps and are symmetrically
arranged about zero energy. In Fig.~\ref{fig1}(a), four edge states
(two-fold degenerate) are present in the regions $-\pi <\theta <-\pi /2$ and
$\pi /2<\theta <\pi $. The four edge states are displayed in Fig.~\ref{fig2}%
(a), wherein the upper panel shows two degenerate edge states of energy $+1$
that are localized on the left and right boundaries, respectively. The
amplitudes are approximately $\{1,1,0\}$ for every three sites from the boundaries. The lower panel shows the two edge states with energy $-1$; the
corresponding amplitudes are approximately $\{1,-1,0\}$ for every three
sites from the boundaries. In Fig.~\ref{fig1}(b), the couplings at the left
and right chain boundaries are the intercell coupling $g_{2}$, and there is
no edge state in any region of $\theta $. In Fig.~\ref{fig1}(c), four edge
states (two-fold degenerate) exist in the region $-\pi /2<\theta <\pi /2$.
In Fig.~\ref{fig1}(d), there are two edge states localized on the left
boundary in region $-\pi /2<\theta <\pi /2$, as shown in Fig.~\ref{fig2}(b),
but no edge states are localized on the right boundary. In Fig.~\ref{fig1}%
(e), there are also two edge states, in the region $-\pi /2<\theta <\pi /2$,
and they are both localized on the right boundary [Fig.~\ref{fig2}(c)]. In
other regions ($-\pi <\theta <-\pi /2$ and $\pi /2<\theta <\pi $), the edge
states are localized on the left boundary [Fig.~\ref{fig2}(b)]. In Fig.~\ref%
{fig1}(f), two edge states exist in the regions $-\pi <\theta <-\pi /2$ and $%
\pi /2<\theta <\pi $, localized on the right boundary.

The existence of edge states is related to the lattice configuration. Edge
states exist when the coupling $g_{1}$ is felt at the lattice boundary. The
configuration with $g_{1}$-$g_{2}$-$g_{1}$- at the boundary results in two
edge states of energy $\pm 1$ in the region $-\pi /2<\theta <\pi /2$,
whereas the configuration with $g_{1}$-$g_{1}$-$g_{2}$- results in two edge
states of energy $\pm 1$ in the regions $-\pi <\theta <-\pi /2$ and $\pi
/2<\theta <\pi $. When the lattice has reflection symmetry, there are four
edge states (two-fold degenerate). The edge states disappear in the
configuration with $g_{2}$-$g_{1}$-$g_{1}$- at the boundary. The edges
states appear at three cases. The edge states plotted in Fig.~\ref{fig2}(a)
are for the lattice with reflection symmetry [Fig.~\ref{fig1}(a)]. Two-fold
degenerate edge states are located at the left and right boundaries,
respectively. In Fig.~\ref{fig2}(b) and~\ref{fig2}(c), the edge states for
the configuration presented in Fig.~\ref{fig1}(e) are plotted; in this
situation, one pair of edge states with energy $\pm 1$ is located on one
boundary. The edge state amplitude is approximately $\left\{ 1,\pm
1,0\right\} $ for every three sites from the boundary. The amplitude decays
according to $(g_{1}/g_{2})^{p}$ in the regions $-\pi <\theta <-\pi /2$ and $%
\pi /2<\theta <\pi $ and according to $\left( g_{2}/g_{1}\right) ^{p}$ in
the region $-\pi /2<\theta <\pi /2$ from the lattice boundary (where $p$ is
the unit cell index). Inside the unit cell, the nonzero amplitudes of edge
states are symmetric (upper panel in Fig.~\ref{fig2}) and antisymmetric
(lower panel in Fig.~\ref{fig2}). For edge state with energy $+1$, the
corresponding amplitude has overall phase difference $e^{i\pi }$ between
neighbor unit cells; for the edge state with energy $-1$, the amplitude has
no such phase difference as illustrated in Fig.~\ref{fig2}.

\section{$\mathcal{PT}$-symmetric non-Hermitian lattice}

\label{III} In this section, we investigate a one-dimensional trimerized
lattice of evanescently coupled optical waveguides. The lattice has three
sublattices $A$, $B$, and $C$, and this spatial arrangement induces
inhomogeneous couplings. The spacing between waveguides $A$ and $B$ is equal
to that between $B$ and $C$ in a unit cell, which differs from the spacing
between waveguides $A$ and its nearest neighbor $C$. This spatial
arrangement induces a periodic modulation in every the third coupling. The
amplitudes of the waveguides in a unit cell are denoted $\psi _{m,A}$, $\psi
_{m,B}$, and $\psi _{m,C}$, where $m$ labels the unit cell index. In coupled
mode theory, the single-mode coupled waveguide lattice is modeled by a
tight-binding system as follows:%
\begin{eqnarray}
i\dot{\psi}_{m,A} &=&-i\gamma _{A}\psi _{m,A}+g_{2}\psi _{m-1,C}+g_{1}\psi
_{m,B}, \\
i\dot{\psi}_{m,B} &=&-i\gamma _{B}\psi _{m,B}+g_{1}\psi _{m,A}+g_{1}\psi
_{m,C}, \\
i\dot{\psi}_{m,C} &=&-i\gamma _{C}\psi _{m,C}+g_{1}\psi _{m,B}+g_{2}\psi
_{m+1,A},
\end{eqnarray}%
where $g_{1}$ is the intracell coupling and $g_{2}$ is the intercell
coupling. The waveguides losses are denoted by $\gamma _{A}$, $\gamma _{B}$,
and $\gamma _{C}$ for sublattices $A$, $B$, and $C$, respectively. Under the
condition $\gamma _{B}-\gamma _{A}=\gamma _{C}-\gamma _{B}\equiv \gamma $,
the system reduces to a $\mathcal{PT}$-symmetric lattice with on-site
potentials $\{i\gamma ,0,-i\gamma \}$ in a three-site unit cell after the
removal of a common loss rate $\gamma _{B}$. The trimerized lattice chain is
described by a $\mathcal{PT}$-symmetric off-diagonal AAH Hamiltonian with
non-Hermitian on-site potentials, i.e., $H=H_{\textrm{AAH}}+(-2i\gamma /\sqrt{3})\sum_{j}\cos \left(
2\pi j/3+\pi /6\right) a_{j}^{\dagger }a_{j}$. The Hamiltonian $H$ can be rewritten in a matrix form, examples of the SSH models are listed in Ref.~\cite{Ruzicka}.

\begin{figure}[tb]
\includegraphics[bb=0 0 435 270, width=7.5 cm, clip]{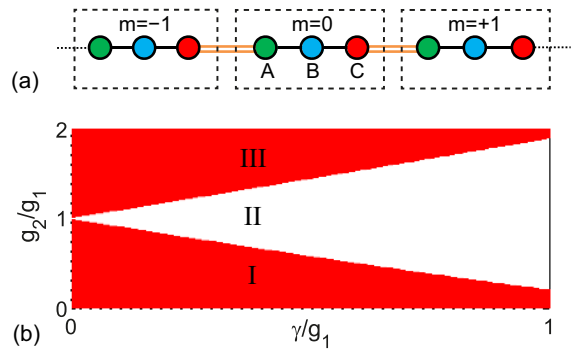}
\caption{(Color online) (a) One-dimensional trimerized $\mathcal{PT}$-symmetric lattice. The unit cell has balanced gain and loss $\{i\gamma,0,-i\gamma\}$ in the dashed rectangles. Gain and loss are represented by green and red, respectively. The passive lattice is displayed in cyan. (b) Phase diagram in the $\gamma$-$g_2$
space. Region II is the broken $\mathcal{PT}$-symmetric phase; region I (III) is the exact $\mathcal{PT}$-symmetric phase in the topologically trivial (nontrivial) region.}
\label{fig3}
\end{figure}

Figure~\ref{fig3}(a) schematically plots the $\mathcal{PT}$-symmetric
lattice under the periodical boundary condition. For $N=3n$, take the Fourier transformation of Bloch waves, the waveguide lattice matrix is expressed as $H=\sum_{k}H_{k}$, where the wave vector $k=2\pi m/n$
($m=1,2,...,n$) and
\begin{equation}
H_{k}=\left(
\begin{array}{ccc}
i\gamma & g_{1} & g_{2}e^{ik} \\
g_{1} & 0 & g_{1} \\
g_{2}e^{-ik} & g_{1} & -i\gamma%
\end{array}%
\right) .
\end{equation}%
Notably, $H_{k}$ is $\mathcal{PT}$-symmetric [$\mathcal{PT}H_{k}(\mathcal{PT)%
}^{-1}=H_{k}$] and includes three energy bands. $\mathcal{P}$ is defined as
the parity operator and satisfies $\mathcal{P}A_{k}\mathcal{P}^{-1}=C_{k}$, $%
\mathcal{P}B_{k}\mathcal{P}^{-1}=B_{k}$ and $\mathcal{P}C_{k}\mathcal{P}%
^{-1}=A_{k}$ (where $A_{k}$, $B_{k}$ and $C_{k}$ are the corresponding
sublattices in Bloch wavevector space). $\mathcal{T}$ is defined as the time
reversal operator, which satisfies $\mathcal{T}i\mathcal{T}^{-1}\mathcal{=}%
-i $. The band gaps are closed at the boundaries of the Brillouin zone: $k=0$ and $%
\pi $. $\mathcal{PT}$ transition occurs at $(\gamma
^{2}-2g_{1}^{2}-g_{2}^{2})^{3}/3^{3}+(2g_{1}^{2}g_{2}\cos k)^{2}/2^{2}=0$.
The phase diagram is depicted in Fig.~\ref{fig3}(b). Region II (white)
represents the region in which $\mathcal{PT}$-symmetry is broken. Regions I
and III are the regions of exact $\mathcal{PT}$-symmetry. Inhomogenity is
necessary for the existence of an exact $\mathcal{PT}$ symmetric phase and
monotonously increases with the degree of non-Hermiticity.

The lattice topology is identified by the Zak phases of the energy bands~%
\cite{Zak,KDing}, which is defined as $\theta =i\int_{-\pi }^{\pi }\mathrm{d}%
k\langle \psi _{R,k}\left\vert \mathrm{d}\psi _{L,k}/\mathrm{d}%
k\right\rangle $ in a non-Hermitian system, where $\psi _{L,k}$ and $\psi
_{R,k}$ are the left and right eigenstates of $H_{k}$~\cite{Esaki,FengLSR}.
The geometric phases accumulated are $\pi ,0,\pi $ for $g_{1}<g_{2}$ when
each Bloch band circles one loop in the Brillouin zone and are $0,0,0$ for $%
g_{1}>g_{2}$. Bulk-edge correspondence means that edge states exist in the
band gaps under an open boundary condition. The exact $\mathcal{PT}$%
-symmetric regions I and III have topologically distinct phases. In the $%
\mathcal{PT}$-symmetric region I, the geometric phases accumulated for the
three bands are all zero, which implies that region I is a topologically trivial phase;
in the $\mathcal{PT}$-symmetric region III, the geometric phases accumulated are
$0$ for the middle band and $\pi $ for the upper and lower bands, which
reflects that region III is a topologically nontrivial phase, and edge
states exist simultaneously between the two band gaps under an open boundary
condition similar to that shown in Fig.~\ref{fig1}. As intercell coupling $%
g_{2}$ is varied from $0$ to $2g_{1}$, the system transitions from an exact $%
\mathcal{PT}$-symmetric phase with trivial topology, through the broken $%
\mathcal{PT}$-symmetric phase, to an exact $\mathcal{PT}$-symmetric phase
with nontrivial topology.

\begin{figure}[tb]
\includegraphics[bb=0 0 280 330, width=8.5 cm, clip]{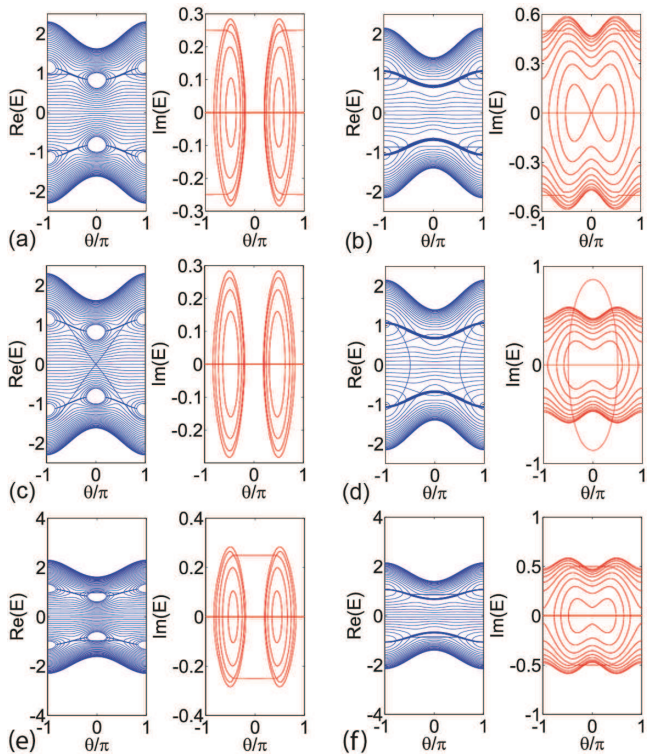}
\caption{(Color online) Energy bands of $\mathcal{PT}$-symmetric
configurations for (a),(b) $N=3n$, (c),(d) $N=3n-1$, and (e),(f) $N=3n-2$. $\protect\gamma=1/2$ for (a),(c),(e), and $\protect\gamma=1$ for (b),(d),(f). $n=30$, $\Delta=1/2$.}
\label{fig4}
\end{figure}

\section{Edge states}

\label{IV} The existence of topologically protected edge state is an
important feature of topological systems. In Fig.~\ref{fig4}, the real and
imaginary parts of the spectra for $N=3n$, $3n-1$, and $3n-2$ at different $%
\gamma $ are plotted. The lattice is the most sensitive to $\mathcal{PT}$%
-symmetric gain and loss at $g_{1}=g_{2}$ ($\theta =\pm \pi /2$). The broken
energy levels are raised as $\gamma $ is increased and the band gaps narrow.
Edge states emerge in the bands but their frequency and amplifying or
damping rate is independent of the intercell coupling $g_{2}$ for a lattice
without defects at the boundary ($N=3n$). For $N=3n-1$, the topology is not
affected by weak non-Hermiticity, and there are no edge states [Fig.~\ref%
{fig4}(c)]. However, an edge state exists at large non-Hermiticity when $%
\gamma >g_{2}$ [Fig.~\ref{fig4}(d)], and its frequency is resonant with that
of the waveguide; the amplifying (damping) rate is $g_{2}$ dependent and
increases with $\gamma $. We calculate the edge states' resonant frequency
and estimate the threshold gain and loss rate as follows.

\begin{figure*}[tb]
\includegraphics[bb=0 0 540 105, width=18.0 cm, clip]{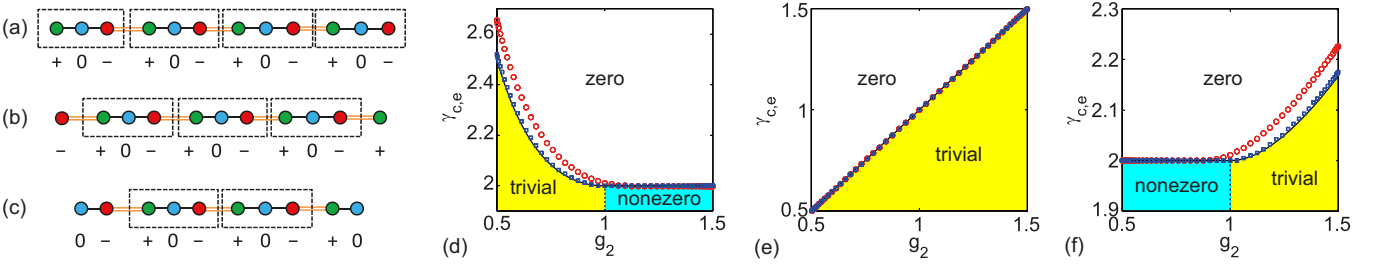}
\caption{(Color online) $\mathcal{PT}$-symmetric lattices with (a) integer
unit cells ($N=3n$); (b) a single additional site at each boundary ($N=3n-1$);
and (c) two additional sites at each boundary ($N=3n-2$). The threshold gain
(loss) rate $\protect\gamma _{\mathrm{c,e}}$ for the real part of the edge
state eigenvalues vanishes, as depicted for (d) $N=3n$, (e) $N=3n-1$, and
(f) $N=3n-2$. Red circles (blue squares) display numerical calculations for
the lattice with $n=30$ ($n=300$); black lines present the analytical
results for infinite systems $n\rightarrow \infty $. Trivial (yellow area),
zero (white area), and nonzero (cyan area) in (d)-(f) indicate the
topologically trivial and nontrivial regions with the real part of the edge
state eigenvalues being zero and nonzero, respectively.}
\label{fig5}
\end{figure*}
Three $\mathcal{PT}$-symmetric configurations are illustrated in Fig.~\ref%
{fig1}(a)-\ref{fig1}(c). Topologically protected edge states exist in the
configurations in Fig.~\ref{fig1}(a) and~\ref{fig1}(c) in the absence of
gain and loss when the system is Hermitian. Notably, the system topology is robust
against the non-Hermiticity. In the presence of gain and loss,
the Hamiltonian still commutes with the $\mathcal{PT}$ operator.
Consequently, the eigenvalues become real or conjugate pairs.
The edge state on one side switches to the other side of the lattice chain
after the $\mathcal{PT}$ operation; therefore, the edge state is not the
eigenstate of the $\mathcal{PT}$ operator and $\mathcal{PT}$ symmetry breaks.
The edge states appear in conjugate pairs, which are robust amplification or
attenuation modes. In the topological non-Hermitian optical lattice,
the coexistence of topology and non-Hermiticity enables the lattice
to function as a robust optical diode.

The Schr\"{o}dinger
equations for the edge states localized on the right side satisfy $\psi
_{N-2}=0$, $E\psi _{N-1}=g_{1}\psi _{N}$, and $\left( E+i\gamma \right) \psi
_{N}=g_{1}\psi _{N-1}$, from which we obtain the edge states eigenvalues
\begin{equation}
E_{-,\pm }=(-i\gamma \pm \sqrt{4g_{1}^{2}-\gamma ^{2}})/2,
\end{equation}%
which are damped modes because the $-i\gamma/2$ term in $E_{-,\pm }$. Indeed, the damping is because that the edge state probability is mainly distributed on the lossy sites. The amplitude of the edge states decays as $%
(g_{1}/g_{2})^{p}$ from the unit cell on the right boundary. The wave
function amplitude is $\{0,\sigma _{+,\pm },1\}$ in each unit cell, where $%
\sigma _{+,\pm }=(i\gamma \pm \sqrt{4g_{1}^{2}-\gamma ^{2}})/(2g_{1})$
indicates the wave function distribution. The edge states can be expressed
as
\begin{equation}
\psi _{r}=[0,(g_{1}/g_{2})^{m-1}\sigma _{+,\pm },(g_{1}/g_{2})^{m-1},\cdots
,0,\sigma _{+,\pm },1]^{T},
\end{equation}%
for system $N=3m$. For edge states localized on the left side, the
eigenvalues are%
\begin{equation}
E_{+,\pm }=(+i\gamma \pm \sqrt{4g_{1}^{2}-\gamma ^{2}})/2,
\end{equation}%
These two edge states are amplified modes. The wave function distribution
is
\begin{equation}
\psi _{l}=[1,\sigma _{-,\pm },0,\cdots
,(g_{1}/g_{2})^{m-1},(g_{1}/g_{2})^{m-1}\sigma _{-,\pm },0]^{T}.
\end{equation}%
The decay factor is $g_{1}/g_{2}$, and the amplitude in the unit cell is $%
\{1,\sigma _{-,\pm },0\}$, where $\sigma _{-,\pm }=(-i\gamma \pm \sqrt{%
4g_{1}^{2}-\gamma ^{2}})/(2g_{1})$. The edge states are depicted in Fig.~\ref%
{fig2}(a) for $\gamma =0$. When the gain and loss are introduced, the
two-fold degenerate edge states in each band gap become one amplified and
one damped pair, with the edge state damping or amplifying rate given by $%
\gamma $. The frequencies of the edge states are reduced from $\pm g_{1}$ to
$\pm \sqrt{g_{1}^{2}-\gamma ^{2}/4}$ in the upper and lower band gaps
separated by $\sqrt{4g_{1}^{2}-\gamma ^{2}}$. The first $\pm $ in the
subscript of $E_{\pm ,\pm }$ indicates that the edge states are damped or
amplified; the second $\pm $ in the subscript of $E_{\pm ,\pm }$ indicates
that the edge states are in the upper or lower band gap.

The edge states for $\gamma =0$ are depicted in Fig.~\ref{fig2}(b) and~\ref%
{fig2}(c). When $\gamma \neq 0$, the right-side edge states are expressed as
\begin{equation}
\psi _{r}=[0,(g_{2}/g_{1})^{m-1}\sigma _{-,\pm },(g_{2}/g_{1})^{m-1},\cdots
,0,\sigma _{-,\pm },1]^{T},
\end{equation}%
and the left-side edge states are expressed as%
\begin{equation}
\psi _{l}=[1,\sigma _{+,\pm },0,\cdots
,(g_{2}/g_{1})^{m-1},(g_{2}/g_{1})^{m-1}\sigma _{+,\pm },0]^{T}.
\end{equation}%
The decay factor is $g_{2}/g_{1}$. $E_{\pm ,+}$ ($E_{\pm ,-}$) corresponds
to the edge states in the upper (lower) band gap. When $g_{2}$ is on the
boundary, two edge states do not exist. If the lattice has a different
boundary of $g_{1}$-$g_{2}$-$g_{1}$- and $g_{1}$-$g_{1}$-$g_{2}$- on the
edge, the system has either right or left edge states, except when $%
g_{1}=g_{2}$.

Three types of $\mathcal{PT}$-symmetric configurations are schematically
illustrated in the $\mathcal{PT}$-symmetric lattice in Fig.~\ref{fig5}. In
all configurations, the repeated unit cells are identical; the lattice
numbers are distinct. The balanced gain and loss in the unit cell are $%
\{-i\gamma ,0,i\gamma \}$. In Fig.~\ref{fig5}(b) and~\ref{fig5}(c), the unit
cells are defective at the edges. This leads to the existence of different
edge states in the corresponding configurations. In particular, edge states
exist in conjugate pairs as the degree of non-Hermiticity increases, and the
real parts of their eigenvalues are zero.

The broken $\mathcal{PT}$-symmetric edge states with real eigenvalues equal
to zero have been demonstrated in honeycomb and square lattices~\cite{Esaki2011}.
In the configuration illustrated in Fig.~\ref{fig1}(b), the chain number is $%
N=3n-1$ and no edge state exists in the Hermitian system (at $\gamma =0$).
Affected by the universal gain and loss, the edge states exist when $\gamma
>g_{2}$. The Schr\"{o}dinger equations for the edge state wave functions
satisfy $\psi _{N-2}=0$, $\left( E+i\gamma \right) \psi _{N-1}=g_{2}\psi
_{N} $, and $\left( E-i\gamma \right) \psi _{N}=g_{2}\psi _{N-1}$. The
eigenvalues of the edge states are $E=\pm i\sqrt{\gamma ^{2}-g_{2}^{2}}$,
implying that the large non-Hermiticity induces a pair of amplified and
damped edge states with their frequency shifted to the waveguide resonant
frequency. Correspondingly, the amplified edge state with energy $E=i\sqrt{%
\gamma ^{2}-g_{2}^{2}}$ is%
\begin{equation}
\psi _{r}=[\cdots ,0,\rho _{-}^{2},\rho _{-},0,\rho _{-},1]^{T},
\end{equation}%
this state is located on the right boundary with decay factor $\rho
_{-}=-i(\gamma -\sqrt{\gamma ^{2}-g_{2}^{2}})/g_{2}$. Conversely, the damped
edge state with energy $E=-i\sqrt{\gamma ^{2}-g_{2}^{2}}$ is%
\begin{equation}
\psi _{l}=[1,\rho _{+},0,\rho _{+},\rho _{+}^{2},0,\cdots ]^{T},
\end{equation}%
this state is located on the left boundary with decay factor $\rho
_{+}=+i(\gamma -\sqrt{\gamma ^{2}-g_{2}^{2}})/g_{2}$. The amplitude of the
unit cell decays as $\rho _{\pm }^{p}$ from the lattice boundary.
\begin{figure}[tb]
\includegraphics[bb=0 0 385 345, width=7.7 cm, clip]{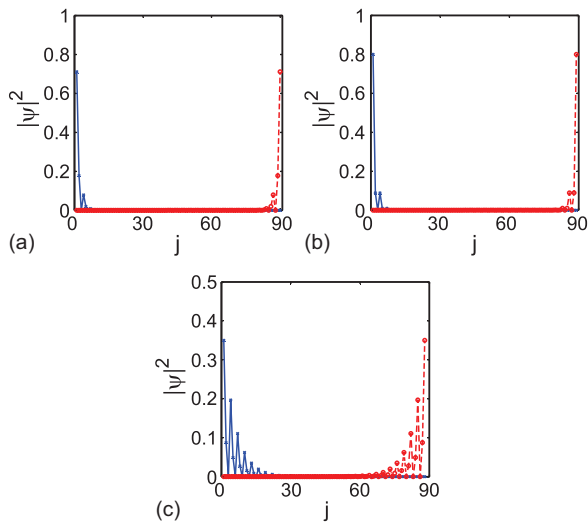}
\caption{(Color online) Edge state probability distributions for
$g_{2}=3/2$ and $\protect\gamma =2.5$. (a) $N=3n=90$, and the energies are
$\pm 2i$; (b) $N=3n-1=89$ and the energies are $\pm 2i$; (c) $N=3n-2=88$ and
the energies are $\pm i/2$.} \label{fig6}
\end{figure}

The edge states are localized at the lattice boundary and decay
exponentially; therefore, we analytically obtain the confinements for the
three configurations as the system size approaches infinity ($N\rightarrow
\infty $). The detailed calculations of the edge states are presented in the
Appendix. For a system constituted by integer unit cells ($N=3n$), the edge
states exist when $g_{2}>g_{1}$ or the gain and loss satisfy
\begin{equation}
\gamma >\gamma _{\mathrm{c,e}}=g_{2}+g_{1}^{2}/g_{2}\text{,}
\end{equation}%
when $g_{2}\leqslant g_{1}$. The lattice with $N=3n-1$ has two defects in
the unit cell at each boundary. Edge states exist when the gain and loss
rates are larger than the coupling $g_{2}$,%
\begin{equation}
\gamma >\gamma _{\mathrm{c,e}}=g_{2},
\end{equation}%
and for a system with $N=3n-2$, which is a lattice with one missing site at
each boundary of the unit cell, the edge states exist when $g_{1}>g_{2}$ or
the gain and loss satisfy%
\begin{equation}
\gamma >\gamma _{\mathrm{c,e}}=g_{2}+g_{1}^{2}/g_{2}\text{,}
\end{equation}%
when $g_{1}\leqslant g_{2}$. The threshold $\gamma _{\mathrm{c,e}}$ is
depicted in Fig.~\ref{fig5} for the three $\mathcal{PT}$-symmetric
configurations. As shown in Fig.~\ref{fig5}(a) and~\ref{fig5}(c), $\gamma _{%
\mathrm{c,e}}$ for finite systems of $n=30$ (red circles) is slightly larger
than that predicted for infinite systems (black lines), whereas $\gamma _{%
\mathrm{c,e}}$ for a finite system of $n=300$ (blue squares) approaches that
predicted for infinite systems. Notably, $\gamma _{\mathrm{c,e}}$ in Fig.~%
\ref{fig5}(b) is system size independent; in this case, there is one pair of
eigenstates $E^{2}=g_{2}^{2}-\gamma ^{2}$ and the amplitudes at the passive
sites (without gain or loss) vanish. Therefore, the wave functions are zero
for every other two sites. The eigenvalues are real and the eigenstates are $%
\mathcal{PT}$-symmetric when $\gamma \leqslant g_{2}$; however, when $\gamma
>g_{2}$, this pair of eigenstates become a pair of edge states with purely
imaginary eigenvalues. The edge states in the three $\mathcal{PT}$-symmetric
configurations for $g_{1}=1$, $g_{2}=1.5$, and $\gamma =2.5$ are plotted in
Fig.~\ref{fig6}. The edge state located on the left boundary in Fig.~\ref%
{fig6}(a) is the amplified mode, which is located on the right boundary in
Fig.~\ref{fig6}(b) and~\ref{fig6}(c). These edge states are related to the
edge state probability and gain (loss) distribution.

\begin{figure}[tb]
\includegraphics[bb=0 0 570 690, width=8.6 cm, clip]{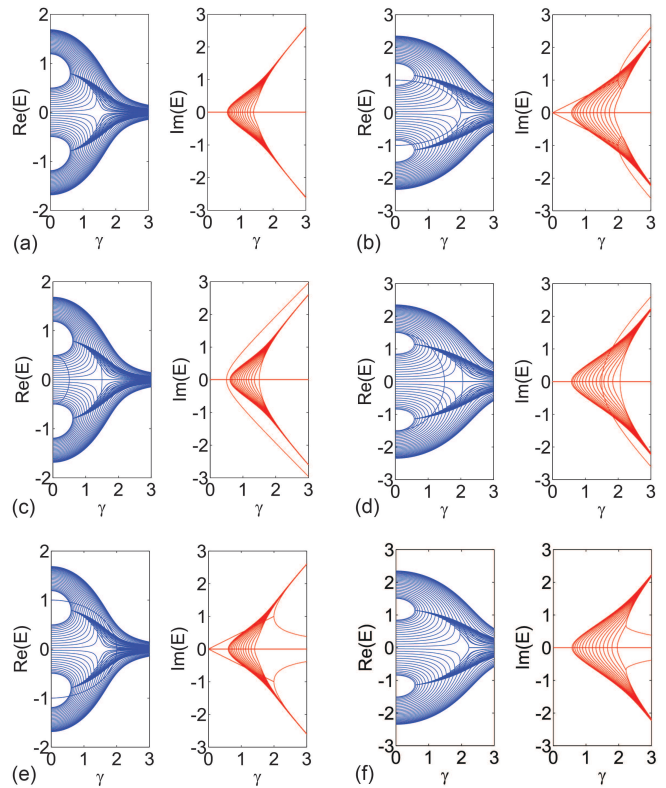}
\caption{(Color online) Energy bands of $\mathcal{PT}$-symmetric
configurations as a function of $\gamma$ for (a),(b) $N=3n$, (c),(d) $N=3n-1$, and (e),(f)
$N=3n-2$. The parameters are $g_2=1/2$ in the left plots and $g_2=3/2$ in the right plots. In all plots, $n=30$.}
\label{fig7}
\end{figure}

The spectra of $\mathcal{PT}$-symmetric configurations are plotted as a
function of $\gamma $ for $g_{1}=1$ and $g_{2}=1/2$ and $3/2$ in Fig.~\ref%
{fig7}. As $\gamma $ is increased, the real parts of the spectra narrow, the
band gap widths are reduced and amplification (attenuation) zero mode edge
states exist when the real parts are decreased to zero at $\gamma _{\mathrm{%
c,e}}$, which is where the edge state frequencies are resonant with the waveguide
lattice and their imaginary parts bifurcate. The thresholds $\gamma _{%
\mathrm{c,e}}$ in Fig.~\ref{fig7} are consistent with the red circles in
Fig.~\ref{fig5}. In Fig.~\ref{fig7}(a), $g_{1}>g_{2}$, and thus the
threshold is dependent on the intercell coupling $g_{2}$, being $\gamma _{%
\mathrm{c,e}}=2.655$; in Fig.~\ref{fig7}(b), $g_{1}<g_{2}$, the threshold is
$2g_{1}$, being $\gamma _{\mathrm{c,e}}=2.0$. Both of these cases are for a
lattice size of $N=3n=90$. The spectra presented in Fig.~\ref{fig7}(c) and~%
\ref{fig7}(d) are those for a lattice size $N=3n-1=89$, where the threshold
equals $g_{2}$, being $\gamma _{\mathrm{c,e}}=0.5$ and $1.5$, respectively.
For a lattice size $N=3n-1$, the band gaps vanish when $g_{1}=g_{2}$. The
bulk states at the top and bottom of the middle band become edge states when
$\gamma >\gamma _{\mathrm{c,e}}$ and $g_{1}>g_{2}$; the edge states are
converted from the bulk states at the bottom of the top band and the top of
the bottom band when $\gamma >\gamma _{\mathrm{c,e}}$ and $g_{1}<g_{2}$.
Figure~\ref{fig7}(e) and~\ref{fig7}(f) displays \ the spectra obtained for a
lattice size $N=3n-2=88$ when $\gamma _{\mathrm{c,e}}=2.0$ and $\gamma _{%
\mathrm{c,e}}=2.226$, respectively. In Fig.~\ref{fig7}(b) and~\ref{fig7}(e),
the edge states exist at any $\gamma $, have broken $\mathcal{PT}$ symmetry,
and have eigenvalues that linearly depend on $\gamma $ until bifurcation
occurs at $\gamma =2.0$. Thereafter, zero mode edge states exist, and
the amplification and attenuation of edge states depend on $(\gamma $, $%
g_{1})$ under a square root function, as indicated by $E_{\pm ,\pm }$. In
Fig.~\ref{fig7}(a),~\ref{fig7}(c),~\ref{fig7}(d),~\ref{fig7}(f), no edge
state exists when $\gamma <\gamma _{\mathrm{c,e}}$, and a pair of bulk
states becomes a conjugate pair of edge states when $\gamma >\gamma _{%
\mathrm{c,e}}$. This pair of edge states is localized on opposite edges with
identical frequency and an equal amount of amplification and attenuation.

\section{Summary}

\label{V}
Topologically nontrivial non-Hermitian trimerized optical lattice is investigated. The $\mathcal{PT}$-symmetric phases, band structure, and topologically protected edge states of a trimerized coupled waveguide lattice with universal non-Hermiticity are demonstrated. This lattice possesses
topologies that are dependent on the degree of non-Hermiticity.
In the topologically nontrivial region, the edge states are related to the configurations
at the lattice boundary and depend on the coupling strengths. Two conjugate edge state pairs exist in the $\mathcal{PT}
$-symmetric configurations. These edge states are symmetrically arranged about
the energy zero. Above a gain (loss) threshold, zero mode edge states exist,
and these states are amplified and damped when propagating in opposite
directions. Asymmetric transport through edge states in the photonic lattice
represents a diode effect. The amplified edge state can be excited by
dynamical creation, which is insensitive to the initial excitation or
lattice imperfections. This robust one-way behavior has potential
applications in optical manipulation, information processing, and
unidirectional lasing.

\acknowledgments We acknowledge the support of NSFC (Grant No. 11605094) and
the Tianjin Natural Science Foundation (Grant No. 16JCYBJC40800).

\appendix*

\section{Solution of the edge states}

In the Appendix, we analyse the appearance of edge states in the three $%
\mathcal{PT}$-symmetric configurations.

For $N=3n$, the Schr\"{o}dinger equations of the edge states satisfy%
\begin{eqnarray}
\left( E-i\gamma \right) \psi _{1} &=&g_{1}\psi _{2}, \\
E\psi _{2} &=&g_{1}\psi _{1}, \\
0 &=&g_{1}\psi _{2}+g_{2}\psi _{4},
\end{eqnarray}%
where $\psi _{j}$ is the wave function at waveguide $j$. We obtain $%
E^{2}-i\gamma E-g_{1}^{2}=0$.

When $\gamma \leqslant 2g_{1}$,
\begin{eqnarray}
E_{\pm } &=&\frac{i\gamma \pm \sqrt{4g_{1}^{2}-\gamma ^{2}}}{2}, \\
\psi _{1} &=&-\frac{g_{2}}{\left( E_{\pm }-i\gamma \right) }\psi _{4}.
\end{eqnarray}%
The wavefunction amplitude yields $\left\vert \psi _{1}\right\vert
^{2}/\left\vert \psi _{4}\right\vert ^{2}=g_{2}^{4}/g_{1}^{2}$, the edge
states exist when $g_{2}>g_{1}$.

When $\gamma >2g_{1}$,%
\begin{eqnarray}
E_{\pm } &=&i\frac{\gamma \pm \sqrt{\gamma ^{2}-4g_{1}^{2}}}{2}, \\
\psi _{1} &=&\frac{2g_{2}}{i(\gamma \mp \sqrt{\gamma ^{2}-4g_{1}^{2}})}\psi
_{4},
\end{eqnarray}%
$\left\vert \psi _{1}\right\vert ^{2}/\left\vert \psi _{4}\right\vert ^{2}>1$
needs $4g_{2}^{2}>(\gamma \mp \sqrt{\gamma ^{2}-4g_{1}^{2}})^{2}$, that is $%
2\left( g_{2}^{2}+g_{1}^{2}\right) -\gamma ^{2}>\mp \gamma \sqrt{\gamma
^{2}-4g_{1}^{2}}$. If $\gamma ^{2}<2\left( g_{2}^{2}+g_{1}^{2}\right) $,
then the wave function amplitude of $E_{+}$ satisfies $\left\vert \psi
_{1}\right\vert ^{2}/\left\vert \psi _{4}\right\vert ^{2}>1$; otherwise, $%
\gamma ^{2}>2\left( g_{2}^{2}+g_{1}^{2}\right) $, the left side is less than
zero and $E_{-}$ satisfies $\left\vert \psi _{1}\right\vert ^{2}/\left\vert
\psi _{4}\right\vert ^{2}>1$, this needs $\gamma ^{2}-2\left(
g_{2}^{2}+g_{1}^{2}\right) <\gamma \sqrt{\gamma ^{2}-4g_{1}^{2}}$, that is
\begin{equation}
\gamma >(g_{2}^{2}+g_{1}^{2})/g_{2}.
\end{equation}%
Notably, $\gamma >2g_{1}$ is already satisfied.

For $N=3n-1$, the Schr\"{o}dinger equations of the edge states satisfy
\begin{eqnarray}
\left( E+i\gamma \right) \psi _{1} &=&g_{2}\psi _{2}, \\
\left( E-i\gamma \right) \psi _{2} &=&g_{2}\psi _{1}, \\
0 &=&g_{1}\psi _{2}+g_{1}\psi _{4}.
\end{eqnarray}%
We obtain $E^{2}=g_{2}^{2}-\gamma ^{2}$, $\psi _{1}=-g_{2}\psi _{4}\left(
E+i\gamma \right) $, $\left\vert \psi _{1}\right\vert ^{2}/\left\vert \psi
_{4}\right\vert ^{2}=1$ when $\gamma <g_{2}$. For $\gamma >g_{2}$, the
energy $E$ becomes pure imaginary
\begin{eqnarray}
E_{\pm } &=&\pm i\sqrt{\gamma ^{2}-g_{2}^{2}}, \\
\psi _{1} &=&-\frac{g_{2}}{(i\gamma \pm i\sqrt{\gamma ^{2}-g_{2}^{2}})}\psi
_{4},
\end{eqnarray}%
$\left\vert \psi _{1}\right\vert ^{2}/\left\vert \psi _{4}\right\vert
^{2}=g_{2}^{2}/(\gamma \pm \sqrt{\gamma ^{2}-g_{2}^{2}})^{2}>1$ for $E_{-}=-i%
\sqrt{\gamma ^{2}-g_{2}^{2}}$; and the corresponding edge state localizes on
the left boundary. The edge state of zero real energy appears at $\gamma
>g_{2}$ for lattice size $N=3n-1$.

For $N=3n-2$, the Schr\"{o}dinger equations of the edge states satisfy%
\begin{eqnarray}
E\psi _{1} &=&g_{1}\psi _{2}, \\
\left( E+i\gamma \right) \psi _{2} &=&g_{1}\psi _{1}, \\
0 &=&g_{2}\psi _{2}+g_{1}\psi _{4},
\end{eqnarray}%
we obtain $E^{2}-i\gamma E-g_{1}^{2}=0$.

When $\gamma \leqslant 2g_{1}$,
\begin{eqnarray}
E_{\pm } &=&\frac{-i\gamma \pm \sqrt{4g_{1}^{2}-\gamma ^{2}}}{2}, \\
\psi _{1} &=&-\frac{g_{1}^{2}/g_{2}}{E_{\pm }}\psi _{4},
\end{eqnarray}%
$\left\vert \psi _{1}\right\vert ^{2}/\left\vert \psi _{4}\right\vert
^{2}=g_{1}^{2}/g_{2}^{2}$, the edge states exist when $g_{1}>g_{2}$.

When $\gamma >2g_{1}$,%
\begin{eqnarray}
E_{\pm } &=&i\frac{-\gamma \pm \sqrt{\gamma ^{2}-4g_{1}^{2}}}{2}, \\
\psi _{1} &=&-\frac{2g_{1}^{2}/g_{2}}{i(-\gamma \pm \sqrt{\gamma
^{2}-4g_{1}^{2}})}\psi _{4},
\end{eqnarray}%
$\left\vert \psi _{1}\right\vert ^{2}/\left\vert \psi _{4}\right\vert ^{2}>1$
needs $4g_{1}^{4}/g_{2}^{2}>(-\gamma \pm \sqrt{\gamma ^{2}-4g_{1}^{2}})^{2}$%
, that is%
\begin{equation}
\left( 2g_{1}^{4}/g_{2}^{2}+2g_{1}^{2}-\gamma ^{2}\right) >\mp \gamma \sqrt{%
\gamma ^{2}-4g_{1}^{2}},
\end{equation}%
If $\left( 2g_{1}^{4}/g_{2}^{2}+2g_{1}^{2}-\gamma ^{2}\right) >0$, that is $%
\gamma ^{2}<2g_{1}^{4}/g_{2}^{2}+2g_{1}^{2}$, then $E_{+}$ satisfies $%
\left\vert \psi _{1}\right\vert ^{2}/\left\vert \psi _{4}\right\vert ^{2}>1$%
; otherwise, $\left( 2g_{1}^{4}/g_{2}^{2}+2g_{1}^{2}-\gamma ^{2}\right) <0$
and needs $\gamma ^{2}-2g_{1}^{4}/g_{2}^{2}-2g_{1}^{2}<\gamma \sqrt{\gamma
^{2}-4g_{1}^{2}}$, that is%
\begin{equation}
\gamma >\left( g_{1}^{2}+g_{2}^{2}\right) /g_{2}.
\end{equation}

\end{document}